\documentstyle[11pt]{article}
\title{On the Limits of Chiral Perturbation Theory}
\author{Thomas Appelquist and John Terning\\
Department of Physics, Yale University, New Haven, CT 06511 \\}
\begin{document}
\setlength{\baselineskip}{24pt}
\maketitle
\vglue 0.8cm
\begin{picture}(0,0)(0,0)
\put(295,250){YCTP-P41-92}
\end{picture}
\vspace{-24pt}
\vglue 0.3cm
\begin{abstract}
We consider the relation between the breakdown scale of chiral perturbation
theory, $\Lambda_\chi$, for large values of $N$ (flavor), and the scale
associated with  ``new" physical thresholds. This question is addressed using
both the linear $\sigma$ model and an asymptotically-free gauge theory to
describe the high energy dynamics. It is suggested  that the massive
physical threshold could be well above $\Lambda_\chi$.
\end{abstract}

\section{ Introduction}

In a recent paper \cite{SS}, Soldate and Sundrum made an interesting
observation about chiral perturbation theory;
especially as applied to technicolor theories with a large
number, $N$, of technifermions.  They showed that
in the large $N$ limit, chiral perturbation theory (a low-energy expansion)
breaks down at a scale no higher than
$\Lambda_\chi \approx 4\pi f/\sqrt{N}$, where $f$ is the
Nambu-Goldstone boson (NGB) decay constant.
Chivukula, Dugan, and Golden \cite{harv}
have pointed out that this could imply the
existence of physical states with masses of order $\Lambda_\chi$ or lower,
a scale
much smaller than $4 \pi f$ for large $N$. If chiral perturbation theory
applies all the way up to $\Lambda_\chi$, then
strong interactions will have set in, and it is therefore difficult to study
this
problem in detail without a strong-coupling computation.  In this paper
we will instead examine a specific, and very simple, model
in order to further elucidate the question of the existence of
low-lying thresholds.  We will then give a more general argument indicating
that there may not
need to be physical states with masses near $\Lambda_\chi$.  We will conclude
with some
remarks on the implications for technicolor models.

\section{ Estimating the Coefficients in a Chiral Lagrangian}

We first discuss methods for estimating
the size of coefficients in an $SU(N)\otimes SU(N)$ chiral Lagrangian.
Recall that, in the
derivative expansion, the leading two-derivative term is:
\begin{equation}
{\cal L}_2 = {f^2 \over 4} {\rm Tr} \partial_\mu U^\dagger \partial^\mu U  ~,
\label{L2}
\end{equation}
where
\begin{equation}
U = \exp\left({{2 i \pi^a T^a}\over{f}}\right) ~,
\label{U}
\end{equation}
and $T^a$ is a generator of $SU(N)$ normalized so that
Tr $\!\!(T^a T^b)={1\over 2} \delta^{ab}$.
Note that ${\cal L}_2$ is
characterized by a single parameter $f$,
giving a dimensionless $\pi \pi$ scattering amplitude proportional to
$p^{2}/f^2$, where $p$ represents the external momenta.
At {\em O}$(p^4)$, there are various terms in ${\cal L}_4$ with a-priori
unknown  coefficients.
One often uses the estimate that the contributions from ${\cal L}_4$
should be the same order of magnitude as the radiative corrections from
NGB loops \cite{Tom,Wein,NDA}. Up to factors of $N$, these corrections will
typically be of
order $p^2/(4{\pi}f)^{2}$ relative to the ${\cal L}_2$ contribution.
(At the one-loop level, this can be corrected by a
logarithm of an ultraviolet cutoff $\Lambda$ over an infrared scale $\mu$.)

To see how this estimate arises, suppose that the natural
scale of the underlying
physics driving the
symmetry breakdown is denoted by $\Lambda_{sb}$.
The ``heavy" physics
above this scale is integrated out, generating an effective
low-energy theory, the chiral Lagrangian, for the degrees of freedom,
the NGB's, relevant
below $\Lambda_{sb}$. (We are primarily interested here in the case of
``dynamical" symmetry breaking, in which the underlying physics is
strongly interacting.  The possibility that the underlying physics is
weakly coupled, as in a weakly coupled linear $\sigma$ model, will
also be considered briefly in section 3.)  All the coupling constants in this
chiral Lagrangian are renormalized at the matching scale
$\Lambda_{sb}$.  Suppose, however, that we want to renormalize at a
somewhat lower
scale $\mu$, by integrating out the Nambu-Goldstone degrees of
freedom between $\Lambda_{sb}$ and $\mu$. Then the coefficients in
${\cal L}_4$ are renormalized by these NGB loops.
Assuming that the contribution
from physics above $\Lambda_{sb}$ does not dominate or cancel the
NGB contribution, this leads to the above estimate (O($p^2/(4\pi f)^2)$)
for the order of magnitude (relative to ${\cal L}_2$)
of the various terms in ${\cal L}_4$.
This simple notion is known as ``naturality" \cite{Tom},
or ``naive dimensional analysis" \cite{Wein,NDA}.

There are two issues that can complicate this simple picture.  First,
the two contributions mentioned above can have a different dependence
on the underlying parameters of the theory.  For example, in QCD, they can
depend
differently on the number of colors, $N_c$.  The NGB
loop computation is
independent of $N_c$ (if $f$ is held fixed), whereas large $N_c$ arguments
\cite{GL,pesktak} suggest that the leading contribution from above
$\Lambda_{sb}$ is proportional to
$N_c$.  This is not surprising since the two, additive, contributions come from
different physics in different energy regimes.   A more subtle problem
is that it may not be possible to calculate the entire pion loop contribution
within the framework of chiral perturbation theory.
This is the case when chiral perturbation theory fails at a scale
$\Lambda_\chi \ll \Lambda_{sb}$.  We know that chiral perturbation theory must
fail when energies of order $\Lambda_{sb}$ are reached since
there
is ``new" physics there that has been
integrated out.  However, there is no guarantee that chiral perturbation theory
converges for the entire range of energies below this scale. In this
paper we will study the relation between the two scales $\Lambda_{sb}$
and $\Lambda_{\chi}$.

\section{ A Linear $\sigma$ Model}

In order to gain a better understanding of what is happening with the chiral
Lagrangian,
it is helpful first to consider a renormalizable model that reduces to the
chiral
Lagrangian when some
of the particles become heavy.  For $N=2$, the simplest such model
is the $O(4)$ linear $\sigma$ model.  This model, when it undergoes
spontaneous symmetry breaking
(the symmetry of the vacuum being $O(3)$),
 reduces to an $O(4)$ nonlinear $\sigma$ model
(which is equivalent to the $SU(2)\otimes SU(2)$ chiral Lagrangian) in the
low-energy
(or $M_\sigma \rightarrow \infty$) limit.  The relation between the
linear and nonlinear theories has been explored in some detail (see ref.
\cite{AB} and further references therein).

The linear $\sigma$ model that reduces to the
$SU(N)\otimes SU(N)$ chiral Lagrangian has been studied by several authors
\cite{Bard,SUN}.  We briefly summarize the properties of
such a model.
Consider the $U(N)_L \otimes U(N)_R$ linear $\sigma$ model:
\begin{eqnarray}
{\cal L}& = &  {1\over 2}\partial_\mu M^\dagger \partial^\mu M
-{{\lambda} \over {4}}\left[{\rm Tr}M^\dagger M M^\dagger M
+{\alpha \over N}\left({\rm Tr}M^\dagger M \right)^2-
2f^2{\rm Tr}M^\dagger M
\right] ~,
\label{Lsig}
\end{eqnarray}
where
\begin{equation}
M = {{\sigma + i \eta^0}\over {\sqrt{N}}} I +
\sqrt{2}(\Sigma^a + i \pi^a )
\,T^a ~.
\label{M}
\end{equation}
$M$ transforms as
\begin{equation}
M \rightarrow L M R^\dagger ~,
\label{LMR}
\end{equation}
where $L$ and $R$ are elements of $U(N)_L$ and $U(N)_R$ respectively.
 The $\Sigma$'s and $\pi$'s are scalars and pseudoscalars
respectively, and transform as adjoints under $SU(N)_V$. The $\sigma$ and
$\eta^0$ are a
scalar and a pseudoscalar respectively, and are singlets under $SU(N)_V$.
The $\eta^0$ and $\pi$'s are massless
NGB's. Note that the
$\eta^0$  is an ``axion", corresponding to the spontaneously
broken $U(1)_A$ symmetry. A term proportional to ${\rm det}
M+ {\rm det} M^\dagger $ could be introduced in the Lagrangian,
which would give a mass to the ``axion", but then the
model would not be
renormalizable for $N > 4$. However, since we are interested mainly in
scattering amplitudes
for the $\pi$'s, the ``axion" will only contribute through loop effects, and
``axion" loops are
suppressed relative to $\pi$ loops by at least one factor of $1/N$.  Thus they
can be safely ignored for large $N$.

At tree level, the condition for minimizing the vacuum energy is:
\begin{equation}
0 = \left.{{\partial V}\over {\partial M_{ij}}}\right|_{M = M_0}=
{\lambda \over 2}\left[M^\dagger_0 M_0 M^\dagger_0 +
\left({\alpha \over N}{\rm Tr}M^\dagger_0 M_0 - f^2\right) M^\dagger_0
\right]_{ji} ~,
\label{VacE}
\end{equation}
where $M_0$ is the vacuum expectation value of $M$ at tree level.
The trivial solution $M_0=0$ is unstable for $f^2>0$.  The other solution is:
\begin{equation}
M_0 M^\dagger_0 = M^\dagger_0 M_0 = f^2 -{\alpha \over N} {\rm Tr}
M^\dagger_0 M_0 ~.
\label{vac}
\end{equation}
Thus we see that $M_0$ is proportional to a unitary matrix, and hence, by a
symmetry transformation,
$M_0$ can be made proportional to the identity, i.e. $M_0 = v I$.  With this
choice of basis, we
have from Eqs.~(\ref{M}) and (\ref{vac}),
\begin{equation}
\sigma_0 = \sqrt{N} v = \sqrt{N \over{1 + \alpha}} \,\,f.
\label{vev}
\end{equation}
So, at tree order, the
$U(N)_L \otimes U(N)_R$ symmetry is spontaneously broken down to $U(N)_V$.
To do perturbation theory in the linear model, we shift the field,
\begin{equation}
M \rightarrow M + v ~.
\label{shift}
\end{equation}
Expanding out the terms in the Lagrangian,
the masses of the $\sigma$ and the $\Sigma$'s at tree level are given by:
$M^2_\sigma = 2 \lambda f^2$, and $M^2_\Sigma =
2 \lambda v^2$.
As promised, the $\eta^0$ and $\pi$'s are massless
NGB's.  For energies much below $M_\sigma$, this
model reduces to the
nonlinear $U(N)\otimes U(N)$ $\sigma$ model, that is, a chiral Lagrangian
supplemented to include the effects of
an ``axion".

Consider this chiral Lagrangian in the large $N$ limit, with $\lambda$ and
$f$ held fixed.  As shown in refs. \cite{SS,harv}, the partial wave $\pi \pi$
scattering amplitude in the singlet channel violates unitarity at energies
of order
$\Lambda_\chi = 4\pi f/\sqrt{N}$, and chiral loop corrections become of order
one at this
same scale.  Thus, chiral perturbation theory breaks down at the
scale $\Lambda_\chi$ (or lower), but how is this scale related to the
lowest non-NGB threshold in
this model?  One
would expect this threshold to be at the $\sigma$ and $\Sigma$
thresholds, at
$\sqrt{2 \lambda}f$ and $\sqrt{2 \lambda}\, v$.  For what might
be thought to be
moderately strong coupling
($\lambda \approx 4\pi^2$), this is much larger
than $\Lambda_\chi$. To understand this difference, it is helpful to return to
the linear model.

We begin with some comments on the loop expansion in the underlying
linear model. First note that diagrams with the
highest powers of $N$ come from graphs with $\pi$ and $\Sigma$ loops.  The
large $N$ analysis for
these diagrams is similar to that for QCD, i.e., since the $\pi$'s and
$\Sigma$'s are in adjoint
representations the counting can be simplified by using the double-line
notation \cite{Cole}.   One
can show that at $L$ loops the leading graphs are of order $N^L$. Thus the loop
expansion parameter is roughly $\lambda N/4\pi^2$, and the expansion can be
expected to break down, defining the onset of strong coupling,
when this parameter becomes of order unity. \footnote{Here $\lambda$ is
defined at scales
of order $M_\sigma$.}

If one demands that the loop expansion parameter in the linear model be of
order unity or smaller, then $M_{\sigma}
= \sqrt{ 2 \lambda}f \leq 2 \sqrt{2} \pi f/\sqrt{N} \approx \Lambda_\chi$.
In this case, the massive physical threshold is of order $\Lambda_\chi$ or
smaller.  At the onset of strong coupling ($\lambda N/4 \pi^2 = O(1)$), the
threshold is of order $\Lambda_\chi$.
It is only in the weak coupling case, $\lambda N/4 \pi^2 \ll 1$,
which is not of central interest here, that the physical threshold will
fall well below $\Lambda_\chi$.

Suppose, next, that $\lambda \gg 4 \pi^2/N$, so that the loop expansion is
useless, and the tree level relation $M_{\sigma}
= \sqrt{ 2 \lambda}f$ is unreliable.  Here, we imagine defining
$\lambda$ at a scale on
the order of the threshold for the production of the massive states of the
theory, whatever they may be.  Chiral
perturbation theory will breakdown at $\Lambda_\chi$, as it must, involving the
same
factors of $N$ that cause
the breakdown of the loop expansion in the
linear model.  Since the
underlying problem is simply the large number of diagrams, however, this
gives us no direct information about finite-mass
singularities of the scattering amplitudes in momentum space, and hence
no direct information about the masses of
the $\sigma$, the $\Sigma$'s, or any other possible massive physical states.
The mass scale of new thresholds cannot be computed in either
chiral perturbation theory or the underlying linear model.

To summarize the discussion of the linear $\sigma$ model as a description of
the high energy dynamics, the massive physical threshold will be below
$\Lambda_{\chi}$ when the loop expansion parameter is small, and it will be
roughly at
$\Lambda_{\chi}$ when this parameter is $O(1)$. The threshold could, however,
be above $\Lambda_{\chi}$ if the loop expansion parameter ( defined at scales
on the order of the threshold) is larger than unity.
In either of the latter two cases, the breakdown
of chiral perturbation theory at $\Lambda_\chi$
is a reflection of the breakdown of the loop expansion.
In the next section we will comment more generally on the possibility
that the massive threshold could be well above $\Lambda_\chi$
by briefly returning to a
more physical discussion of the low-energy chiral Lagrangian.

\section{ Beyond the Linear $\sigma$ Model and Beyond the Loop Expansion}

As pointed out in ref. \cite{harv}, the most general $\pi^a \pi^b
\rightarrow \pi^c
\pi^d $ scattering amplitude for
the $SU(N) \otimes SU(N)$ chiral Lagrangian can be written as:
\begin{eqnarray}
a(s,t,u)^{a,b;c,d}&= &\delta^{ab}\delta^{cd}A(s,t,u) + d^{abe}d^{cde}
B(s,t,u) \nonumber\\
& & + (s \leftrightarrow t,
b \leftrightarrow c) + (s \leftrightarrow u, b \leftrightarrow d) ~,
\label{scatt}
\end{eqnarray}
where $d^{abc}$ is defined by $\{T^a,T^b\} = d^{abc}T^c + \delta^{ab}/N$.

Now if we choose any functions for $A(s,t,u)$ and $B(s,t,u)$ that are
consistent with
chiral symmetry and unitarity, there should be a
chiral Lagrangian (with an infinite number of terms) that reproduces our
arbitrary
scattering amplitude when an exact calculation is performed \cite{Wein}.
The constraint that chiral symmetry imposes on the two functions is that
the leading
low energy behavior (as $s$, $t$, and $u \rightarrow 0$) agrees with the
tree order
result \cite{harv} derived from Eq.~(\ref{L2}):
\begin{eqnarray}
A(s,t,u)&=&{ {2 s}\over{N f^2}}+ . . .\nonumber \\
B(s,t,u)&=& {s \over f^2}+ . . .
\label{chiralconst}
\end{eqnarray}

Suppose that $A$ and $B$ are functions whose only singularities
are cuts
corresponding to multi-$\pi$ thresholds, i.e., there are no resonances.
There is some
chiral Lagrangian, consisting of Eq.~(\ref{L2}) and higher-order terms,
that can produce such a
scattering amplitude.  Of course, chiral perturbation theory, at any finite
order, will be useless for
energies above $\Lambda_\chi = 4\pi f/\sqrt{N}$, since the expansion
parameter $p^2 / \Lambda_\chi^2$ is
larger than one.  However, when the chiral perturbation series is summed
to all orders,
the functions $A$ and $B$ can be obtained, even for $p^2 \gg \Lambda_\chi^2$.
This could still be well below
the scale of ``new" physics.  This is reminiscent
of the fact
that some series can converge, even when the expansion parameter is larger
than one, e.g.,
the Taylor expansion of $(p^{2}/f^{2})e^{-p^2/\Lambda^2_\chi
\ln(-a p^2/\Lambda^2_\chi)}$
 (where $a$ is some constant)
when the exponential is expanded
in powers of $p^2/\Lambda^2_\chi \ln(-a p^2/\Lambda^2_\chi)$.

This type of behavior, with no massive physical threshold at
all, is, of course, extreme.  It would require much
cancellation in the chiral loop expansion in order to obtain the requisite
inverse factorials.  However, between the two extremes of thresholds
at $\Lambda_\chi$, and no thresholds, there lies a range of possibilities for
the threshold mass.
Reality may lie somewhere in this range.
To say this in another way, the actual form of the scattering amplitude is
almost surely not as simple as the above exponential form.
In particular, terms of order $(p^{2})^n$ in the low energy expansion will
naturally come multiplied by various powers of $\ln(-p^{2}/\Lambda_\chi^{2})$,
i.e. $\ln^{m}(-p^{2}/\Lambda_\chi^{2})$, where $m=0,1,...n-1$. Even so,
the fact that $\Lambda_\chi$ may set the scale for all the terms
in chiral perturbation theory does not necessarily mean that there must be a
massive physical threshold at $\Lambda_{\chi}$.

\section{ Implications for Technicolor}

The
motivation for studying theories with large numbers of flavors
comes from technicolor model building.
There, for example, a popular class of
models has one family of technifermions, and hence an $SU(8)\otimes SU(8)$
flavor symmetry.
In technicolor the question of the how $\Lambda_\chi = 4\pi f/\sqrt{N}$
is related to the scale of ``new" physics can be rephrased as:
what is the relation between $f$ and
the technicolor
confinement scale? (The confinement scale $\Lambda_{tc}$,
is roughly the scale of
``new" physics in technicolor theories, and $f$ is related to the weak
scale by $f = 250$ GeV $\sqrt{2/N}$.)
A typical procedure for estimating $f$ relies on
the approximation that
$f$ is determined by diagrams with only one fermion loop. This is reliable in
large-$N_c$ QCD, and is
also assumed in Pagels-Stokar type analyses.  These analyses
lead to the rough estimate $4 \pi f \approx \sqrt{N_{tc}} \Lambda_{tc}$,
where $N_{tc}$ is the number of technicolors. This would place
$\Lambda_{\chi}$ well below $\Lambda_{tc}$ for $N$ well above $N_{tc}$.

 With $N$ large, however, these one-fermion-loop
analyses are not likely to be reliable. Additional technifermion loops enter
with a factor of $N$, and could be at least as important as the gauge
corrections to the single fermion loop. Even with $N$ constrained to preserve
the asymptotic freedom of the theory (as in the above one-family model with
$N_{tc}$ even as low as $2$), it could be that contributions with arbitrary
numbers of fermion loops are equally important. Among all these contributions,
there will appear, of course, the low energy loops of NGB's, that can be dealt
with
(chiral)-perturbatively below $\Lambda_\chi$, but give no direct information
about the relation of $4{\pi}f$ to the confinement scale $\Lambda_{tc}$. The
contributions from momenta above $\Lambda_{tc}$, on the other hand, do provide
this connection, but may
involve an $N$ dependence that renders the above Pagels-Stokar estimate
invalid. Whether the correct result lowers the scale of ``new" physics toward
$\Lambda_{\chi}$ or leads to an even larger gap between the two scales is not
clear to us.

In order to get a more physical picture of what may be going on in
such theories, imagine integrating out the ``heavy" physics
in an asymptotically-free  technicolor theory.
Since a low-energy expansion must be performed in order to obtain the
chiral Lagrangian, the mass
that sets the scale for $f$ (and hence
$\Lambda_\chi$) in the low-energy theory is the mass (actually the
zero-momentum mass)
of the ``lightest" non-NGB, e.g. the techni-$\rho$.  For
particles with widths of the same order as their masses (as would
be expected, for example, for a techni-$\rho$ in a theory
with $N$ as large as 8), there is
no simple connection between the zero-momentum mass and the physical
mass.  Thus, it is reasonable to expect that there may not be a
direct connection between $\Lambda_\chi$ and the location of the
lowest resonance peak, i.e. the scale of ``new" physics.

The estimates for $S$ and $T$ in technicolor models suffer from the
same uncertainties as the estimates of $f$,
since all current methods for estimating $S$ and $T$
\cite{pesktak,SandT}
rely either on scaling QCD (where the quark loop expansion presumably does
work)
up to technicolor (where the technifermion loop expansion may or may not
work), or on an
explicit single technifermion loop computation.  In particular, estimates of
$S$ based on the techni-$\rho$ mass may be unreliable since there is no
simple connection between $f$ and the physical techni-$\rho$ mass in large $N$
technicolor models.

\section{ Conclusions}

We have suggested that the fact that chiral perturbation theory for the
$SU(N)$ $\otimes
SU(N)$ chiral Lagrangian breaks down at very low energies ($\approx 4\pi
f/\sqrt{N}$) for large $N$ does not necessarily
indicate unexpected,
low-lying, physical states.  We have noted that the potential breakdown of the
technifermion loop expansion (for large $N$) for the estimation of $f$, $S$,
and $T$ in terms of the parameters of a technicolor theory, reflects
the difficulty of relating the chiral symmetry breakdown scale to the
threshold for ``new" physics.

\newpage
\noindent \medskip\centerline{\bf Acknowledgments}
\vskip 0.15 truein
We thank S. Chivukula and R. Sundrum for helpful
discussions. JT acknowledges the financial support of a Superconducting
Super Collider
fellowship from the Texas National Research Laboratory Commission.
This work was supported in part by the U.S.  Department of Energy
under contracts DE-AC-02-76ERO3075 and DE-FG-02-84ER40153.

\end{document}